\documentclass[superscriptaddress,twocolumn,showpacs,prl,floatfix]{revtex4}

\usepackage{color}
\usepackage{tabularx}
\usepackage{epsfig}
\usepackage{amsmath}
\usepackage{graphicx}

\begin{document}

\title{The Fortuin-Kasteleyn and Damage Spreading transitions in
Random bond Ising lattices}

\author{P. H.~Lundow} \affiliation {Department of Theoretical Physics,
  Kungliga Tekniska h\"ogskolan, SE-106 91 Stockholm, Sweden}

\author{I. A.~Campbell}
\affiliation{Laboratoire Charles Coulomb,
  Universit\'e Montpellier II, 34095 Montpellier, France}

\begin{abstract}
The Fortuin-Kasteleyn and heat-bath damage spreading temperatures
$T_{FK}(p)$ and $T_{ds}(p)$ are studied on random bond Ising models of
dimension two to five and as functions of the ferromagnetic
interaction probability $p$; the conjecture that $T_{ds}(p) \sim
T_{FK}(p)$ is tested. It follows from a statement by Nishimori that in
any such system exact coordinates can be given for the intersection
point between the Fortuin-Kasteleyn $T_{FK}(p)$ transition line and
the Nishimori line, $[p_{NL,FK},T_{NL,FK}]$. There are no finite
size corrections for this intersection point. In dimension three, at the
intersection concentration $[p_{NL,FK}]$ the damage spreading
$T_{ds}(p)$ is found to be equal to $T_{FK}(p)$ to within $0.1\%$. For
the other dimensions however $T_{ds}(p)$ is observed to be
systematically a few percent lower than $T_{FK}(p)$.

\end{abstract}


\maketitle

\section{Introduction}
The physics of Ising spin glasses (ISGs) in the "ordered" regime below
the freezing temperature $T_g$ has been intensively investigated for
decades; the paramagnetic regime above $T_g$ has attracted less
attention.  However, in addition to the standard ordering (or freezing)
transitions other significant "critical" temperatures within the
paramagnetic regime can be defined operationally and estimated
numerically with high precision.

We have studied random-bond Ising models (RBIM) in dimensions two,
three, four and five having ferromagnetic near neighbor interactions
with probability $p$ and antiferromagnetic interactions with
probability $1-p$ over the whole range of $p$ in the paramagnetic
regime.  The Hamiltonian is $H = \sum_{ij}-J_{ij}S_{i}S_{j}$ where the
sum is taken over all nearest neighbour bonds $ij$.

For the discrete $\pm J$ interaction distribution the bond
values are thus chosen according to
\begin{equation}
P(J_{ij}) = p\delta(J_{ij} - J) + (1 - p)\delta(J_{ij} + J)
\label{PJij}
\end{equation}
We will set $J=1$ and use when convenient inverse temperatures $\beta
= 1/T$.  The pure Ising model is recovered for $p = 1$ and the pure
antiferromagnetic model for $p = 0$, while the standard strong
disorder bimodal spin glass case corresponds to $p = 1/2$.

In addition to the standard ferromagnetic and spin glass ordering
transitions, on the RBIM phase diagram in the paramagnetic regime
other physically significant lines can be operationally defined as
functions of $p$. In addition to the Griffiths line ($T_{gr}(p) =
T_c$) \cite{griffiths:69,matsuda:08} and the Nishimori line
\cite{nishimori:81} there is the Fortuin-Kasteleyn (FK) transition
line $T_{FK}(p)$
\cite{fortuin:72,coniglio:80,coniglio:91,dearcangelis:91,cataudella:91,zhang:93}
and the heat-bath damage spreading transition line $T_{ds}(p)$
\cite{derrida:87,derrida:89,dearcangelis:89,campbell:91}. Early large
scale relaxation measurements on the ISG in dimension three were
interpreted \cite{ogielski:85} in terms of a dynamic transition at the
Griffiths temperature (i.e. $T_{G}(p) = T_{c}$), with a qualitative
change in relaxation behavior of the autocorrelation function from
non-exponential to exponential.

It can alternatively be considered that $T_{FK}$ or $T_{ds}$ defines a
dynamic transition.  At $T_{FK}$ the FK cluster size diverges so the
standard cluster flipping algorithms \cite{swendsen:87,wolff:89} break
down.  At $T_{ds}$, the time scale for the "coupling from the past"
equilibration criterion $D(t) \to 0$ diverges at large $L$
\cite{propp:96,bernard:10} so this criterion similarly breaks down. \emph{A
  priori} it seems plausible that the two dynamic breakdowns should
occur at similar, if perhaps not identical, temperatures as has been
conjectured by a number of authors
\cite{dearcangelis:91,campbell:94,yamaguchi:10}.

In the following discussion it is shown that there are universal
analytic relations for the intersection point between $T_{FK}(p)$ and
the Nishimori line \cite{nishimori:81}. With these exact values in
hand together with the present accurate numerical damage spreading
data a critical test can be made of the conjectured rule
$T_{ds}(p)\sim T_{FK}(p)$. This is shown to be a good approximation,
almost exact in dimension three and accurate to a few percent in
dimensions two, four and five. However the equality between the
temperatures is not a general rule.

\section{Transition definitions}

%First, consider the Nishimori line (NL).

Nishimori \cite{nishimori:81} has shown for the RBIM that, due to
extra symmetries of the problem, a number of quantities may be
computed exactly when the equality
\begin{equation}
(1 - p_{NL})/p_{NL} = \exp(-2\beta_{NL})
\label{NLdef}
\end{equation}
holds; this condition defines the Nishimori line (NL), a line
traversing the entire $\beta(p)$ phase diagram including both
paramagnetic and ferromagnetic regimes. In particular on the NL the
internal energy of the system per bond (or edge) is
\begin{equation}
U_{NL} = -[2p_{NL} - 1]   = - \tanh(\beta_{NL})
\label{NLU}
\end{equation}
This energy per bond has exactly the value that independent bonds
would have at the same temperature. We will see below that indeed on
the NL the positions of the satisfied bonds are uncorrelated
\cite{nishimori:86,nishimori} and that bond positions remain very
close to random over a wide strong disorder regime around $p=1/2$.  In
other words anywhere on the NL the satisfied bonds are uncorrelated on
average, so \emph{a fortiori} the FK active bonds are distributed at
random.

The rule for the Fortuin-Kasteleyn (FK) transition line
\cite{fortuin:72,coniglio:80,coniglio:91,dearcangelis:91,cataudella:91,zhang:93},
corresponding initially to the pure ferromagnets, is to first select
for some particular equilibrium configuration the entire set of
"satisfied" bonds $ij$ where $J_{ij}S_{i}S_{j}$ is positive. These
bonds are then decimated at random leaving a fraction $[1 -
  \exp(-2/T)]$ of "active" satisfied bonds. The FK stochastic
transition at $T_{FK}$ occurs when the set of active bonds percolates
through the lattice. From the way in which the bonds are laid down, in
the general case this is correlated bond percolation, as opposed to
the standard uncorrelated bond percolation with random bond
occupation. Remarkably, for pure ferromagnets it can be proved that
the FK transition $\beta_{FK}(p=1)$ coincides exactly with the Curie
temperature $\beta_{c}(p=1)$ \cite{fortuin:72}. The Swendsen-Wang and
Wolff cluster algorithms \cite{swendsen:87,wolff:89} speed up
equilibration dramatically as long as the FK clusters remain of finite
size.

The same operational definition can also be used elsewhere in the
phase diagram; in particular $T_{FK}(p=1/2)$ was estimated numerically
for strong disorder RBIM on square, cubic, and triangular lattices
\cite{coniglio:91,dearcangelis:91,cataudella:91,zhang:93}. It turned
out that $T_{FK}(p=1/2)$ was in each case much higher than the
standard spin glass ordering temperature $T_{g}(p=1/2)$. Even in the
two-dimensional case where the glass temperature is zero,
$T_{FK}(p=1/2)$ is similar to but lower than the Griffiths temperature
$T_{gr}=T_c(p=1)$ \cite{cataudella:91,zhang:93}. The physical basis
for the percolation transition temperature was explained in terms of
"FK droplets" for the case of the fully frustrated lattice and general
Potts $q$ in Refs.~\cite{prakash:93,cataudella:93}. Imaoka {\it et
  al.} \cite{imaoka:97} estimated $T_{FK}(p)$ numerically over the
entire range of $p$ for the square and triangular RBIM lattices.

Below we will introduce a further temperature $T_{FKr}(p)$ closely
related to $T_{FK}(p)$.  By construction the fraction of satisfied
bonds $n_{s}(p,T)$ in equilibrium is related to the equilibrium
internal energy per bond $U(p,T)$ through $U(p,T) = 1-2n_{s}(p,T)$ so
the fraction of all bonds which are FK active bonds is just
\cite{campbell:94}
\begin{equation}
  P_{a}(p,T) =(1-U(p,T))[1-\exp(-2/T)]/2.
  \label{Pa}
\end{equation}
$U(p,T)$ and hence $P_{a}(p,T)$ can be readily measured numerically to
high precision.  It was noted \cite{campbell:94} that if the satisfied
bond positions are assumed to be uncorrelated, the FK transition would
occur at a temperature such that $P_{a}(p,T_{FK}) \equiv P_{c}$ where
$P_{c}$ is the standard random bond percolation concentration for the
lattice.

We will refer to the temperature where the condition
\begin{equation}
  P_{a}(p,T_{FKr}) = P_{c}
  \label{PaPc}
\end{equation}
is satisfied as $T_{FKr}(p)$ ($r$ standing for random). This
conjecture led to estimates for $T_{FKr}(p=1/2)$ on square and on
cubic lattices in higher dimensions \cite{campbell:94} which were in
good agreement with the numerical estimates for $T_{FK}(p=1/2)$
available at the time.

It can be noticed that for the pure Ising square lattice ferromagnet
at criticality $U(\beta_{c})= -1/2^{1/2}= -0.7071\ldots$ while
$-\tanh(\beta_{c})=-0.4121\ldots$ which is very different because the
positions of the satisfied bonds are strongly correlated. However the
critical FK concentration of active bonds is $P_{a}(p=1,T_c)=1/2$
which is "accidentally" equal to $P_{c}=1/2$ for this lattice. For the
pure Ising ferromagnet on the triangular lattice at criticality on the
other hand $P_{a}(p=1,T_c)=0.352208\ldots$ which is not quite equal to
$P_{c}=0.347296\ldots$ for this lattice.

Finally, a heat-bath damage spreading transition, $T_{ds}(p)$ can be
defined
\cite{derrida:87,derrida:89,dearcangelis:89,campbell:91}. Heat-bath
update rules are applied at fixed temperature to two initially
non-identical spin configurations $A(0)$ and $B(0)$ (which are not
necessarily equilibrated) of a given sample (meaning $A$ and $B$ have
exactly the same sets of interactions $J_{ij}$); the random number
used in each subsequent single spin update step is the same for both
configurations. At high enough temperatures, on annealing under this
procedure for sufficient time $t$, $A(t)$ and $B(t)$ will become
identical; below the damage temperature $T_{ds}$ the "damage" ($D(t)$
the Hamming distance between $A(t)$ and $B(t)$ divided by $L^d$) will
stabilize for long times at a temperature dependent non-zero
value. For the pure ferromagnet case $T_{ds}(p=1)$ is equal to the
Curie temperature $T_{c}(p=1)$ \cite{derrida:87,derrida:89}. For the
strong disorder spin glass, just as $T_{FK}(p=1/2)$ is much higher
than the glass temperature $T_{g}(p=1/2)$, $T_{ds}(p=1/2)$ is also
much higher than $T_g(p=1/2)$
\cite{derrida:87,derrida:89,dearcangelis:89,campbell:91}. The damage
spreading transition for a given coupling algorithm, in particular
heat bath (HB), can be described as a "regular to chaotic" dynamic
transition from the viewpoint of the "coupling from the past"
approach. When the damage falls to zero after a sufficient anneal time
it is a guarantee that the system has been strictly equilibrated
\cite{chanal:08,chanal:10,bernard:10}. In systems with frustration
this guarantee breaks down when $T < T_{ds}(p)$, so perfect
equilibration in this sense for the regime near the critical
temperature cannot be achieved.

It should be noted that $T_{ds}$ depends on the updating protocol,
with for instance Glauber updating giving very different results from
HB updating. It turns out that even within the HB protocol the precise
value obtained for $T_{ds}$ changes slightly depending on whether
sequential or random updating is used. The results reported here are
for random updating. With sequential updating the observed damage
spreading temperature is of the order of $1\%$ lower.

Early comparisons of numerical data from different groups indicated
that $T_{ds}(p=1/2) \sim T_{FK}(p=1/2)$ for dimensions two and three
\cite{dearcangelis:91,campbell:94}, and on this basis it has been
conjectured by a number of authors
\cite{dearcangelis:91,campbell:94,yamaguchi:10} that $T_{ds}(p) \sim
T_{FK}(p)$ is a general rule defining a joint dynamic transition
temperature above which relaxation is exponential in the long time
limit and below which relaxation is chaotic. It should be noted that
in practice it is very hard to identify such a transition directly
from autocorrelation function decay $q(t)$ data.

Recently Yamaguchi \cite{yamaguchi:10} focused attention on behavior
on the NL. He provided conjectures suggesting that the intersection of
the $T_{FK}(p)$ line and the NL would occur when
\begin{equation}
  p_{NL,FK}=(1+P_{c})/2
  \label{pYg}
\end{equation}
and
\begin{equation}
  \beta_{NL,FK} =\ln[(1+P_{c})/(1-P_{c})]/2
  \label{TYg}
\end{equation}
where again $P_{c}$ is the random percolation concentration for the
lattice. It turns out that because of the analytic value for the
energy $U_{NL}(p)$ on the NL, Eq.~\eqref{NLU}, the Yamaguchi condition is
strictly equivalent to the equality $T_{NL,FK}= T_{NL,FKr}$,
meaning that this condition holds if the FK active bonds are
distributed at random. The random bond conjecture for $\beta_{FK}$,
Eq. (4) of Ref.~\cite{campbell:94}, is identical to Eq. (3.3) of
Ref.~\cite{yamaguchi:10}.

In fact Nishimori \cite{nishimori:86} years earlier had made the
strict statement : "We have also proved independence of the local
internal energy of different bonds, which indicates that the system
effectively splits into uncorrelated sets of bonds on the [NL] in the
phase diagram." In other words anywhere on the NL the satisfied bonds
are uncorrelated \emph{on average}, so \emph{a fortiori} the FK active
bonds are distributed at random. Hence at the NL-FK intersection
concentration $p_{NL,FK}$, we have $T_{FK}\equiv T_{FKr}(p)$, meaning
that the conditions for this intersection conjectured by Yamaguchi,
Eq.~\eqref{pYg} and Eq.~\eqref{TYg}, hold exactly for any RBIM lattice
if the FK transition is a random active bond percolation
transition. Remarkably, a further consequence of Nishimori's statement
is that this equality should hold \emph{for the mean over many
  samples} independently of the sample size $L$.

The numerical results below show that for dimension two, and
presumably in other dimensions also, in addition to this identity at
$p_{NL,FK}$, for a wide range of $p$ around $p=1/2$, $T_{FK}(p) \sim
T_{FKr}(p)$ remains a very good approximation, as was conjectured for
$p=1/2$ in Ref.~\cite{campbell:94}. The active bonds are thus
essentially uncorrelated at $\beta_{FK}(p)$ in this wide "high
disorder" regime.  It should be noted however that as a general rule
the positions of active bonds are correlated so $T_{FK}(p)$ is not
equal to $T_{FKr}(p)$.

Yamaguchi \cite{yamaguchi:10}, following
Refs.~\cite{dearcangelis:91,campbell:94} for the case of $p=1/2$,
also conjectured that $T_{ds}(NL) = T_{NL,FK}$ at the intersection.
The present measurements show that in dimension two $T_{ds}(p)$ is in
fact systematically lower than $T_{FK}(p)$ by a few $\%$ over the
entire range of $p$, except at and near the pure ferromagnet limit
$p=1$. At $p_{NL,FK}$ for cubic models in dimensions three to five
the conjecture holds to within about $0.1\%$ in dimension three, to
$2\%$ in dimension four and to $4\%$ in dimension five. The general
relation $T_{ds}(p)\sim T_{FK}(p)$ is therefore simply a reasonably
good approximation. In dimensions two, three and four the present data
show that the damage spreading transition temperature like the FK
transition temperature is very insensitive to $p$ for a wide range of
$p$ around $p=1/2$.

\section{FK Numerical results in dimension two}

The two dimensional lattices studied were the square lattice and the
triangular lattice.  On both lattices the standard phase diagrams as
functions of $p$ are well established. The pure ferromagnetic Curie
temperatures $T_{c}(p=1)$ are exactly $T_{c}(sq) = -2/\ln[2^{1/2} -1]
= 2.26918 \ldots$ and $T_{c}(tri) = 4/\ln 3 = 3.64096 \ldots$
respectively. When $p$ is lowered from $p=1$, the Curie temperature
$T_{c}(p)$ drops gradually until a critical point is reached at
$p_{r}$ on the Nishimori line where the Curie temperature tends
suddenly to zero with weak re-entrant behavior
\cite{nishimori:81,nobre:01,wang:03,amuroso:04,parisen:09,thomas:11}. Between
the re-entrant regime and $p = 1/2$ the system can be considered a
"spin glass" but with no finite temperature ordering transition. The
square lattice phase diagram for $p < 1/2$ is the exact mirror image
of the phase diagram for $p > 1/2$ with antiferromagnetic order taking
the place of ferromagnetic order as the lattice is bipartite.  For the
triangular lattice on the other hand there is no finite temperature
order below $p=1/2$ as the fully frustrated antiferromagnetic limit at
$p=0$ is approached.

The internal energy per bond $U(p,T,L)$ at equilibrium was estimated
numerically.  For small triangular lattices, $3\le L \le 11$, the
energy as a function of temperature for each specific sample was
calculated exactly (up to numerical precision) using transfer
matrices. Once a sample is picked we split the lattice into $L$ parts,
each part corresponding to a transfer matrix $A_i$. After choosing a
numerical value of $T$ we evaluate the matrices and compute the trace
of their product to obtain
$Z(p,T,L)=\mathrm{tr}\left(A_1\,A_2\cdots,A_L\right)$. This standard
approach is described in great detail for the Ising case in section 3
of \cite{lundow:02} and is of course easily adapted to the spin glass
case.  We thus computed $\langle \ln Z(p,T,L) \rangle$ for
$p=0,0.02,0.04,\ldots,0.50$ and $p=0.51,\ldots,0.98,0.99,1$ and some
70-80 values of $T$. The number of samples ranged between 16384 for
$L=3,4,5$ and then down to only 1024 for $L=11$. Taking the average
$\partial \ln Z/\partial \beta$ then provides us with $\langle
U(p,T,L) \rangle$. Values at $(p,T)$ outside the computed data grid
were obtained through 3rd order interpolation.

\begin{figure}
  \includegraphics[width=3.5in]{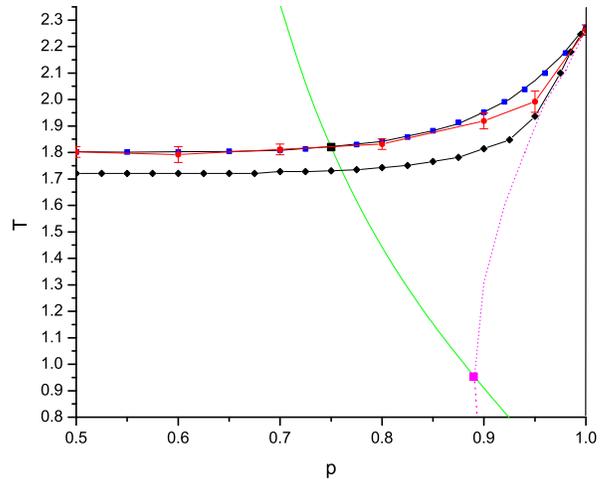}
  \caption{(Color online) The square lattice transitions, temperatures
    $T$ as functions of ferromagnetic interaction concentrations
    $p$. The pink dashed line on the right indicates the ferromagnetic Curie
    temperatures, data taken from Ref.~\cite{thomas:11}. The green
    solid traversal line is the exact Nishimori line (NL). Red circles : FK
    transitions $T_{FK}$, Ref.~\cite{imaoka:97}. Blue squares :
    random active bond percolation line $T_{FKr}$. Black diamonds :
    heat bath damage spreading transition $T_{ds}$. The errors
    on $T_{FK}$ and $T_{ds}$ are of the size of the points. Large black square  % CHECK THIS
    : exact intersection point $T_{NL,FK}$. } \protect\label{fig:1}
\end{figure}

\begin{figure}
  \includegraphics[width=3.5in]{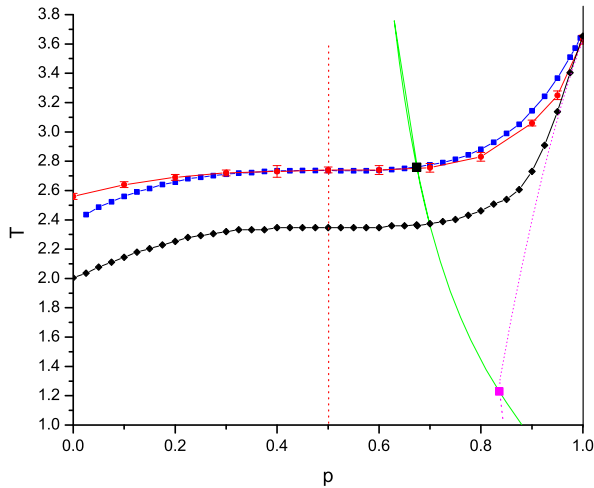}
  \caption{(Color online) The triangular lattice transitions,
    temperatures $T$ as functions of ferromagnetic interaction
    concentrations $p$. The pink dashed line on the right indicates
    the ferromagnetic Curie temperatures. The green solid traversal
    line is the exact Nishimori line (NL). The red dashed vertical
    line in the centre incates $p=0.5$. Red circles : FK transitions
    $T_{FK}$, Ref.~\cite{imaoka:97}. Blue squares : random active bond
    percolation line $T_{FKr}$. Black diamonds : heat bath damage
    spreading transition $T_{ds}$. The errors on $T_{FK}$ and $T_{ds}$
    are of the size of the points.  Large black square : exact
    intersection point $T_{NL,FK}$. } \protect\label{fig:2}
\end{figure}

For larger $L$ (for the square and triangular lattice) the Monte Carlo
data were collected after equilibration using standard Metropolis
updating. At each $(p,T)$, and for each sample, we collected at least
a few million measurements of $U$, and of course even more for
moderate $L$. For the square lattices we used $L=16,32,64$ on 64
different samples with $0.5\le p < 1$ in steps of $0.025$.  For the
triangular lattice we used $L=16,32$ on 32 different samples but with
$0<p<1$ in steps of $0.025$ (slightly denser at high and low
$p$). Some 50 values of $T$ were used for both lattices. Again,
intermediate values of $\langle U\rangle$ were obtained by
interpolating the values at the $(p,T)$ grid points.  By interpolation
of points derived from Eq.~\eqref{Pa} and the consistency condition
Eq.~\eqref{PaPc} the temperature $T_{FKr}(p)$ where the fraction of FK
active bonds $P_{a}(p,T)$ is equal to $P_c$ can be estimated to high
precision.  In Figures~\ref{fig:1} and \ref{fig:2} the $T_{FKr}(p)$
values are compared to the directly measured $T_{FK}(p)$ values
\cite{cataudella:91,imaoka:97} for the square and triangular lattices
respectively. The damage spreading temperatures to be discussed later
are also shown. 

The error in the $T_{FKr}$-estimates is found by first solving
Equation~\eqref{Pa} for each individual sample (using the
interpolation we mentioned above) which gives us a sample-to-sample
standard deviation from which we obtain the standard error of the
average-sample $T_{FKr}$. Once the standard error is established we
are free to plot a smoothed version (from fitting a polynomial of high
degree) of $T_{FKr}$ versus $p$. Needless to say we have not estimated
the error in $U$, and hence $T_{FKr}$, for each individual sample, but
we simply assume this error should be reflected in the
sample-to-sample variation. 

In dimension two, $P_{c} = 1/2$ for the square lattice and $P_{c} =
2\sin(\pi/18)=0.347296355\ldots$ for the triangular lattice. The exact
NL-FK intersection values from Eqs.~\eqref{pYg} and \eqref{TYg} are
then : $p_{NL,FK} = 3/4$, $\beta_{NL,FK} = \ln(3)/2= 0.5493063\ldots$
for the square lattice, and $p_{NL,FK} = 0.6736 \ldots $,
$\beta_{NL,FK} = 0.362371\ldots$ for the triangular lattice.  The
estimated $T_{FK}(p)$ and $T_{FKr}(p)$ lines run directly through
these exact NL-FK intersection points as they should,
Figures~\ref{fig:1} and \ref{fig:2}.

\begin{figure}
  \includegraphics[width=3.5in]{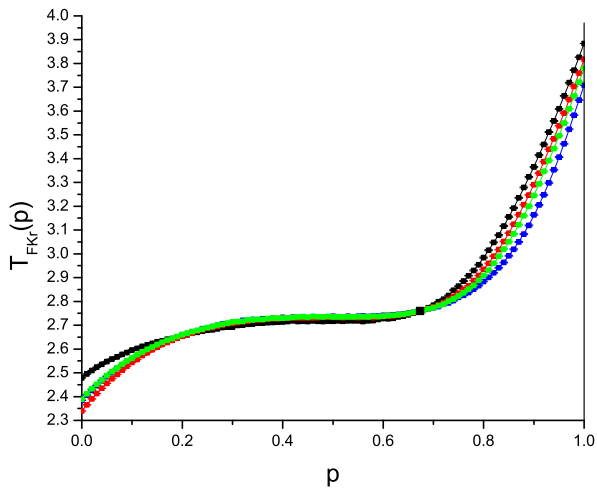}
  \caption{(Color online) The size effect on triangular lattice
    temperatures $T_{FKr}$ at which the fraction of FK sites is equal
    to the random site percolation concentration for the lattice
    $P_{c}$, Eq.~\eqref{PaPc}. Temperatures $T_{FKr}$ as functions of
    ferromagnetic interaction concentrations $p$ for lattice sizes $L$
    from 3 to 11. $L=3, 4, 5, 11$ black, red, green, blue; from top to
    bottom on the right hand side and from bottom to top on the left
    hand side. The black square is the exact $T_{NL,FK}$ intersection
    Eqs.~\eqref{pYg} and \eqref{TYg}.} \protect\label{fig:3}
\end{figure}

\begin{figure}
  \includegraphics[width=3.5in]{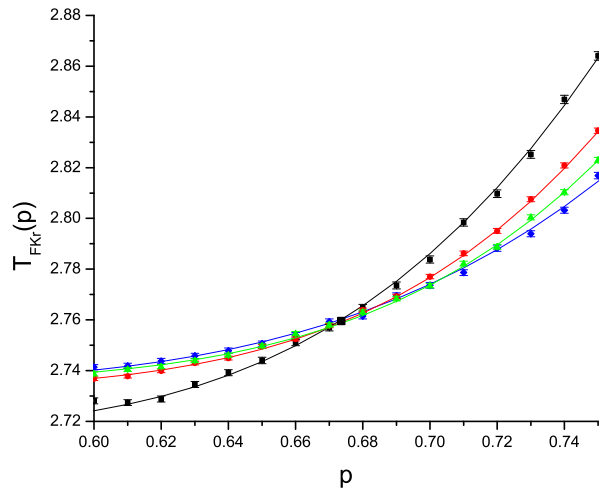}
  \caption{(Color online) As for Fig.~\ref{fig:3}, a closeup of the
    region around the $T_{NL,FK}$ intersection (square).  $L=3, 4, 5,
    11$ (black square, red circle, green triangle, blue diamond; from
    top to bottom on the right hand side and from bottom to top on the
    left hand side).  The error bars are the standard error when
    measuring $T_{FKr}$ after solving Eqs.~\eqref{Pa} and \eqref{PaPc}
    and correspond to 16384 samples for $L=3,4,5$ and 1024 for
    $L=11$. The curves are 10th degree polynomials fitted to the whole
    range from $p=0$ to $p=1$.  }\protect\label{fig:4}
\end{figure}

The measured equilibrium energy per bond is slightly higher than the
random bond energy $-\tanh(\beta)$ for $\beta > \beta_{FKr}$ and
slightly lower for $\beta < \beta_{FKr}$.

To within the high numerical precision the estimated
$\beta_{FKr}(p_{NL,FK},L)$ at the NL-FK intersection point is
independent of $L$ down to $L=3$, Figures~\ref{fig:3} and
~\ref{fig:4}.  The observed absence of finite scaling corrections at
the NL also follows from Nishimori's general statement quoted above
\cite{nishimori:86}. For $p$-values to the left and right of
$p_{NL,FK}$ there are weak finite size effects of opposite signs.
This absence of mean finite size scaling deviations arises because
$T_{FKr}$ depends only on the energy. Other parameters can still show
finite size scaling deviations \cite{hasenbusch:08}.  The large size
limits for $T_{FK}(p)$ and $T_{FKr}(p)$ remain equal to each other to
within the numerical precision of the $T_{FK}(p)$ points
\cite{imaoka:97} over the strong disorder range of $p$ extending from
$p=1/2$ to $p_{NL,FK}$ in the square lattice, and from $p \sim 0.3$ to
$p_{NL,FK}$ in the triangular lattice. Over these ranges of $p$ the
satisfied bonds at $T_{FK}$ are very close to being uncorrelated; for
instance at $p=1/2$ on the square lattice
$-U(p,\beta_{FKr})/\tanh(\beta_{FKr}) = 0.9709$ which remains near to
the uncorrelated value of $1$. The further randomness introduced by
the FK decimation is sufficient to render the active bond positions
essentially uncorrelated.  The $T_{FK}(p)$ \cite{imaoka:97}(or
$T_{FKr}(p)$ which has been measured here with higher precision) is
rather insensitive to $p$ within this range, so that the measured
$T_{FK}(1/2)$ is similar to the exact $T_{NL,FK}$ on both
lattices. For both lattices in the more strongly ferromagnetic ranges
$p > p_{NL,FK}$, and in the range $p < 0.3$ near to the fully
frustrated $p=0$ limit in the triangular lattice, $T_{FK}(p) >
T_{FKr}(p)$. At these concentrations the satisfied bonds are
significantly correlated at $T_{FK}(p)$ and the FK decimation does not
sufficiently compensate the correlations so as to produce randomness
among the active bonds.

%The $p_{NL,FK}$ intersection point is special in having strictly
%uncorrelated satisfied bonds and so uncorrelated percolation for the
%FK bonds. It should be noted that it is not necessary to have
%strictly uncorrelated satisfied bonds in order to have a critical FK
%bond percolation fraction equal to $P_c$. A counter example is
%provided by the critical ferromagnetic square lattice at $p=1$, where
%the satisfied bonds are strongly correlated but nevertheless the
%critical FK bond fraction is again exactly equal to $P_c = 1/2$. This
%is not so for the triangular lattice at $p=1$.

\section{Damage spreading numerical results in dimension two}

\begin{figure}
  \includegraphics[width=3.5in]{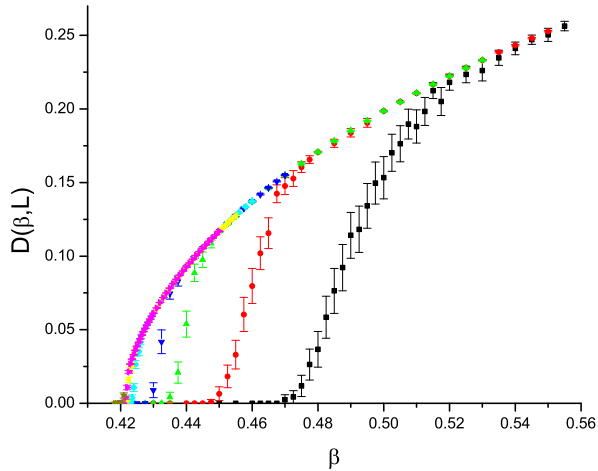}
  \caption{(Color online) The damage spreading $D(\beta,L)$ for the
    triangular lattice at $p=0.6736$. The lattice sizes are $L=48, 64,
    96, 128, 256, 384, 512, 1024$ (black square, red circle, green
    triangle, blue inverted triangle, cyan diamond, yellow left
    triangle, pink right triangle, brown star. Right to left in the
    region where the curves separate). Shown error bars are the
    standard error. The exact FK transition inverse temperature is
    $\beta = 0.362371 \ldots$ so clearly distinct from the limiting
    damage spreading inverse temperature.}  \protect\label{fig:5}
\end{figure}

For dimension two the critical temperatures $T_{ds}(p)$ below which
the long time large $L$ heat bath damage spreading $D(p,T,t)$ tends to
a non-zero value are also shown in Figures~\ref{fig:1} and
\ref{fig:2}. The values given are obtained from extrapolating data at
increasing sizes, Figures~\ref{fig:5} and \ref{fig:6}, to infinite
size. On the basis of data for square and cubic lattices at $p=1/2$ it
was conjectured earlier that $T_{FK}(p) \sim T_{ds}(p)$
\cite{dearcangelis:91,campbell:94,yamaguchi:10} as is the case in the
pure ferromagnet $p=1$ limit. However the present data demonstrate
that while the curves for $T_{ds}(p)$ and $T_{FK}(p)$ lie near
together and are of very similar shape, in particular both being
almost independent of $p$ for the range near $p=1/2$, it is clear that
$T_{ds}(p) < T_{FK}(p)$ except when $p$ tends to the ferromagnetic
limit $p=1$. So unfortunately it is not possible to define a unique
joint "dynamic transition temperature". We will see that this is true
also in higher dimensions.

The present accurate critical value for the strong disorder square
lattice $T_{ds}(1/2)= 1.69(2)$ is in good agreement with the earlier
value $T_{ds}(1/2)= 1.70$ \cite{campbell:91} and is very close to the
"regular to chaotic" dynamic transition temperature estimated for the
same lattice in Ref.~\cite{bernard:10} from the divergence of the
coupling time with $L^2$ , where the data indicate a transition at
$T_{ds}(1/2) \sim 1.72$.

\section{Dimensions three, four and five}

It is numerically much more demanding to estimate $T_{FK}(p)$
precisely through direct measurements (as in \cite{imaoka:97}), in
particular allowing for finite size effects, than to estimate
$T_{ds}(p)$ to the same level of accuracy.  For cubic lattices in
dimensions three, four and five the random bond critical
concentrations $P_{c}$ though not exact have been estimated to very
high precision, $P_{c}(3)=0.2488126(5)$, $P_{c}(4)=0.1601310(10)$ and
$P_{c}(5)=0.11811718(3)$ respectively
\cite{lorenz:98,dammer:04}. Using the exact NL-FK intersection point
expressions, Eqs.~\eqref{pYg} and \eqref{TYg}, one thus has
$p_{NL,FK}= 0.6244$ and $\beta_{NL,FK}= 0.2541466$ in dimension three,
$p_{NL,FK}= 0.58006$ and $\beta_{NL,FK}= 0.161521$ in dimension four,
and $p_{NL,FK}= 0.55906$ and $\beta_{NL,FK}= 0.118671$ in dimension
five. Hence as $p_{NL,FK}$ and $\beta_{FK}(p_{NL,FK})$ can be taken as
known almost exactly, it is sufficient to estimate
$\beta_{ds}(p_{NL,FK})$ numerically (allowing carefully for finite
size corrections) to obtain an accurate estimate of the ratio
$T_{ds}(p_{NL,FK})/T_{FK}(p_{NL,FK})$ for each dimension.

\begin{figure}
  \includegraphics[width=3.5in]{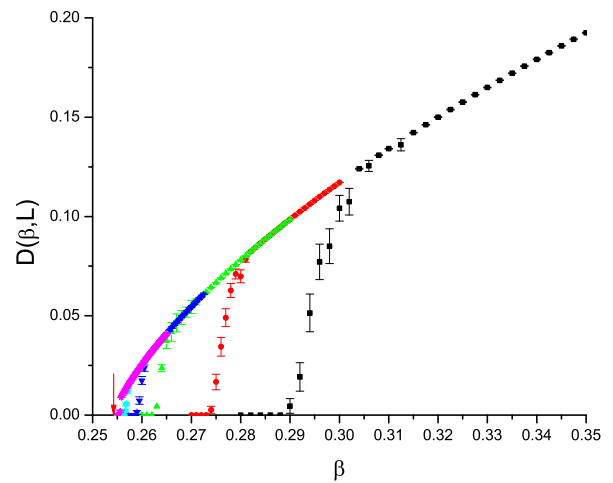}
  \caption{(Color online) The cubic lattice equilibrium damage
    spreading $D(\beta,L)$ as a function of size and inverse
    temperature for $p=0.6244$. Sizes : $L=12, 16, 24, 32, 48, 64$
    (black square, red circle, blue triangle, green inverted triangle,
    cyan diamond, pink star ; right to left in the region where the
    curves separate). Shown error bars are the standard error. The
    extrapolated intersection with the $\beta$ axis gives the infinite
    size critical $\beta_{ds}$. In this case $\beta_{ds}=0.25432(15)$
    can hardly be distinguished from the exact FK transition inverse
    temperature $\beta_{FK}= 0.25414 \ldots$, red arrow.  }
  \protect\label{fig:6}
\end{figure}

The equilibrium damage $D(L,\beta)$ was measured for given $L$ as a
function of $\beta$, and the results were extrapolated to obtain an
estimate of the $\beta_{ds}$ value at which $D(\infty,\beta)$ falls to
zero, Figures~\ref{fig:5} and \ref{fig:6}.  There is a clear envelope
curve for all $L$ in each case, with the data points leaving the curve
later and later as $L$ increases. There are various ways to
extrapolate to infinite $L$. One efficient method is to fit the
envelope points assuming that near $\beta_{ds}$ the behavior follows
$D(\beta)=A(\beta-\beta_{ds})^B$ and adjusting $B$ to obtain a
straight line. We have also used a simple scaling formula
$\beta_{ds}(L)=\beta_{ds}+C\,L^{-\lambda}$ to verify. Of course, a
certain statistical uncertainty will enter depending on which $L$ to
include in the fitting process which we take into account in the final
error estimate.

The infinite $L$ damage spreading temperatures $T_{ds}(p_{NL,FK})$
were estimated for the central values of $p_{NL,FK}$ in the three
dimensions : $T_{ds}(p_{NL,FK}) = 3.932(2)$ in dimension three,
$T_{ds}(p_{NL,FK}) =6.057(10)$ in dimension four, and
$T_{ds}(p_{NL,FK}) =8.13(1)$ in dimension five.  The observed ratios
$T_{FK}(p_{NL,FK})/T_{ds}(p_{NL,FK})$ are equal to $1.0005(5),
1.022(2)$ and $1.036(10)$ respectively. In dimension three
$T_{ds}(p_{NL,FK})$ is indistinguishable from $T_{FK}(p_{NL,FK})$,
while in dimensions four and five the values appear tantalizingly
close but not identical. We have no explanation for the striking
similarity between the two temperatures for the particular case of
dimension three.  Adding a little more detail for this particular
case, the individual $\beta_{ds}(L)$ were estimated to an accuracy of
$\pm 0.0001$ for $L=48,64$, $\pm 0.00025$ for $L=32$, $\pm 0.0005$ for
$L=16,20,24$ and $\pm 0.001$ for $L=12$. Using the exponent
$\lambda=2.05$ gives a projected $\beta_{ds}$ that depends only to a
very small degree on which $L$ are included in the fit (though we
always include $L=48,64$).  We receive a median value of
$\beta_{ds}=0.25432$ and a standard deviation of $0.00015$ giving us
$T_{ds}=3.932(2)$.  Similar methods were used to estimate $T_{ds}$ for
the other dimensions.

\begin{figure}
  \includegraphics[width=3.5in]{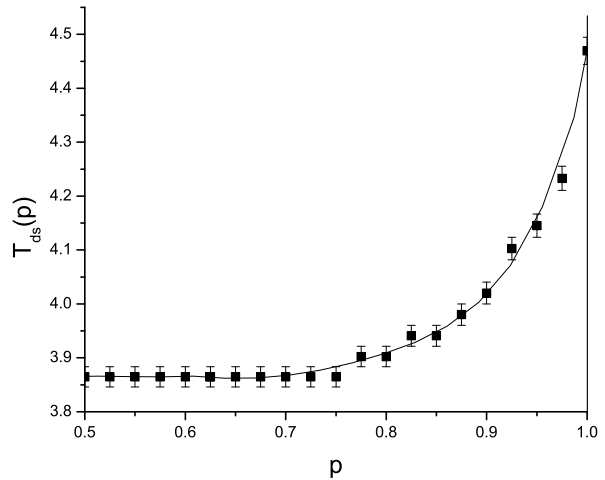}
  \caption{ The damage spreading critical temperature $T_{ds}(p)$ for
    a 3d cubic lattice size $L=32$. The error bars correspond mainly
    to the residual uncertainty in the extrapolation to the
    intersection with the $\beta$ axis.}  \protect\label{fig:7}
\end{figure}

\begin{figure}
  \includegraphics[width=3.5in]{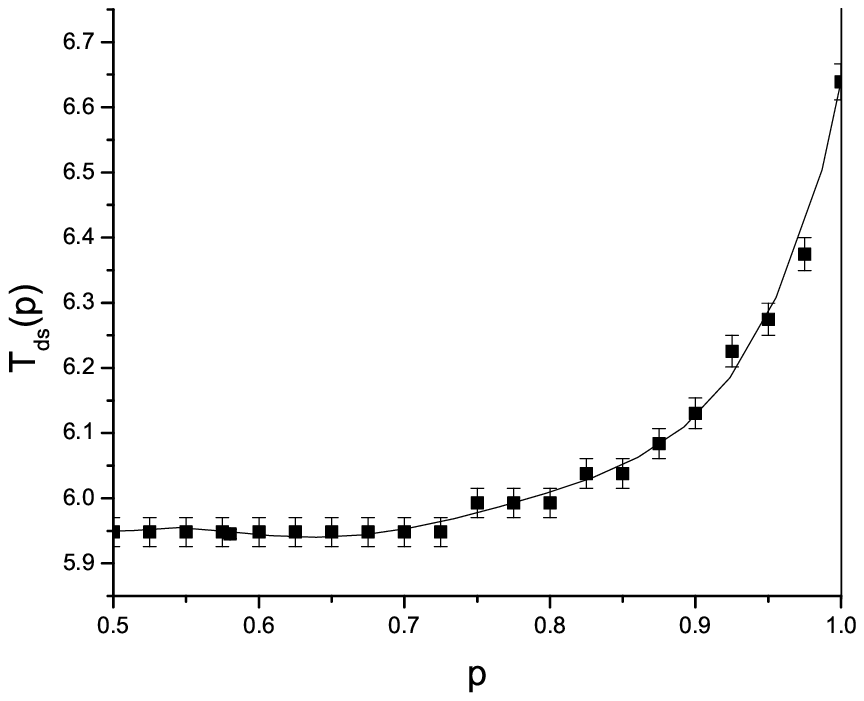}
  \caption{ The damage spreading critical temperature $T_{ds}(p)$ for
    a 4d cubic lattice size $L=16$.  The error bars correspond mainly
    to the residual uncertainty in the extrapolation to the
    intersection with the $\beta$ axis.}  \protect\label{fig:8}
\end{figure}

As in dimension two, the curve for $T_{ds}(p)$ is almost flat over a
wide range of $p$ around $p=1/2$ for dimensions three and four,
Figures~\ref{fig:7} and \ref{fig:8} (the case of dimension five was
not studied). These data are for fixed $L$ ($L=32$ and $L=16$
respectively) so the $T_{ds}(p)$ values are not quite equivalent to
the values quoted for $T_{ds}$ at $p_{NL,FK}$.

Judging from the behavior observed in dimension two, it seems very
plausible to assume that the ratio $T_{FK}(p)/T_{ds}(p)$ is always
practically independent of $p$ in the strong disorder regime $1/2 < p
< p_{NL,FK}$.

It can be noted that numerical data for the time dependence of the
autocorrelation function $q(t)$ at all temperatures in two dimensional
fully frustrated systems have been interpreted in terms of exponential
relaxation with logarithmic corrections due to vortex-vortex
interactions \cite{walter:11}. The results indicate that in these
systems (which have well established FK and damage transitions) there
is no dynamic critical temperature in the Ogielski sense
\cite{ogielski:85}.

%In particular in dimension two, $P_{c} = 1/2$ for the square lattice
%and $P_{c} = 2\sin(\pi/18)$ for the triangular lattice.

\section{Damage clusters}

It is of interest to examine in some detail the damage clusters in
dimension two where they can be readily visualized.  Keeping
$\beta>\beta_{ds}$ and letting $t\to\infty$ (or at least allowing for
equilibration) how does the damage actually spread?  Note that we
define a cluster as a connected component in the lattice induced by
the damaged sites, i.e. a maximal set of damaged sites such that there
is a lattice path (made up of horizontal and vertical steps) between
each pair.  The sum of the individual cluster sizes is thus the damage
without normalization, i.e. $L^d\,D(\beta)$. We have collected data
on cluster sizes for $d=2,3,4,5$. For $d=2$ we have only used the
square lattice, not the triangular lattice.

\begin{figure}
  \includegraphics[width=3.5in]{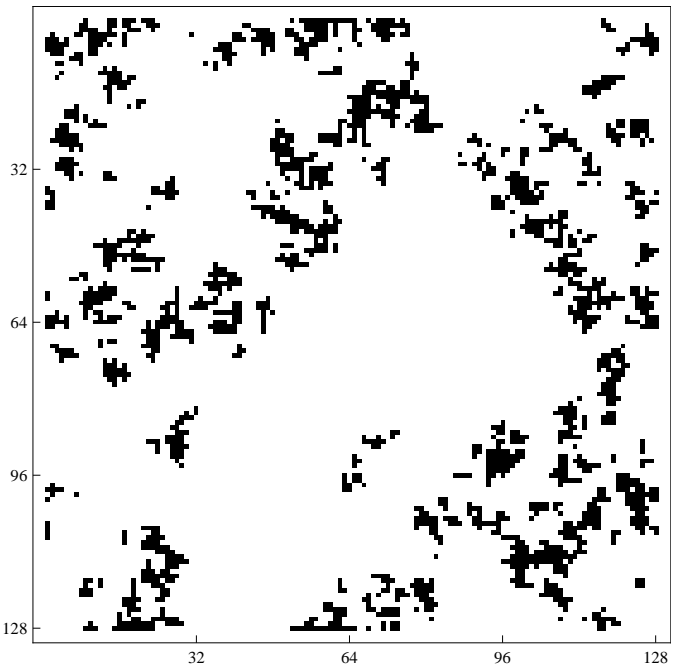}
  \caption{An illustrative snapshot of an instantaneous damage site
    configuration for an $L=128$ square lattice at $p=0.75$ and
    temperature $T= 1.61$, just below the damage spreading
    transition.} \protect\label{fig:snap}
\end{figure}

What we see in a square lattice snapshot at $\beta$ near $\beta_{ds}$
and at any arbitrary fixed time is not, as in the case for percolation
(including FK percolation) or random graphs near $p_c$, the formation
of a giant cluster, but rather a large number of very small
clusters. At subsequent times the clusters evolve and flutter through
the lattice. An example is shown in Fig.~\ref{fig:snap}

The measurements take place as above for the damage spreading;
starting from random (infinite temperature) spin configurations, at
each time step we update $L^d$ randomly selected sites, then search
for all clusters and their sizes. On a measurement we collect the
number of clusters, the average cluster size, the size of the largest
cluster and the size of a randomly selected cluster. As before, we
collect $10^6$ measurements from each sample after discarding between
$50000$ and $250000$ time steps (depending on $L$) to allow for
equilibration of the damage spreading. We then average over the time
steps to get the average for a particular sample. The data are then
averaged over the samples; we have only used eight samples in each
case for the cluster measurements (for the damage speading we used
between 8 and 128 samples depending on lattice size). In all cases we
have set $p=p_{NL,FK}$.

\begin{figure}
  \includegraphics[width=3.5in]{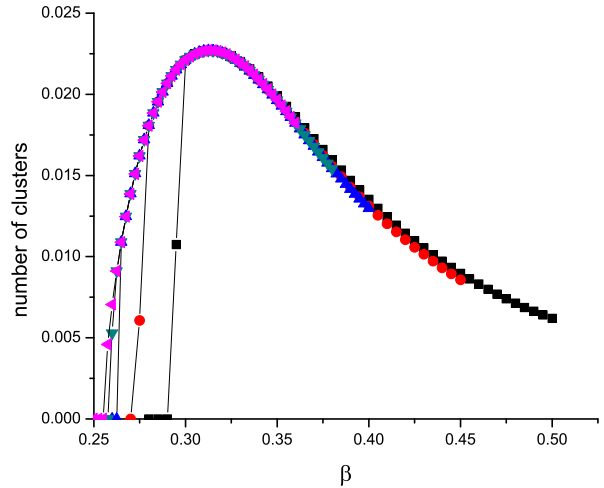}
  \caption{(Color online) The number of clusters $n_{c}(\beta,L)$ on
    the 3d cubic lattice normalized by $L^3$ at $p=p_{NL,FK}$. Sizes
    $L= 12, 16, 24, 32, 48$ shown as black squares, red circles, blue
    triangles, green inverted triangles, pink left triangles. The
    errors are smaller than the size of the
    points.}\protect\label{fig:10}
\end{figure}

First we discuss the expected number of clusters, which we denote
$n_{c}(\beta,L)$.  In Figure~\ref{fig:10} we plot $n_{c}(\beta,L)/L^3$
versus $\beta$ for the simple cubic lattice. Equivalent behavior is
found also for $d=2,4,5$.  Note that for the square lattice all
lattice sizes agree on a global maximum probability located at 
$\beta_{\max}=0.656(1)$, where $n_{c}(\beta_{\max},L)\sim
0.0130(1)\,L^2$.  For $d=3$ we obtain $\beta_{\max}=0.313(1)$ and
$n_{c}(\beta_{\max},L)\sim 0.0227(1)\,L^3$, for $d=4$ we get
$\beta_{\max}=0.211(1)$ and $n_{c}(\beta_{\max},L)\sim
0.0200(1)\,L^4$, while $d=5$ gave $\beta_{\max}=0.177(1)$ and
$n_{c}(\beta_{\max},L)\sim 0.0308(1)\,L^5$. Thus the damage is always
distributed over $O(L^d)$ clusters.

\begin{figure}
  \includegraphics[width=3.5in]{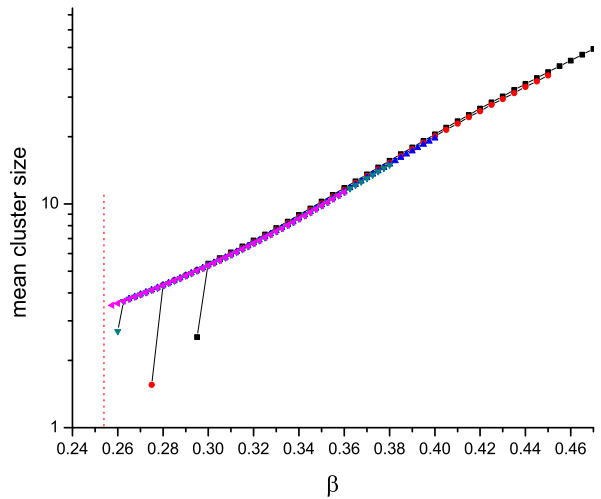}
  \caption{(Color online) The mean cluster size $s(\beta,L)$ on the 3d
    cubic lattice at $p=p_{NL,FK}$. Sizes $L= 12, 16, 24, 32, 48$
    shown as black squares, red circles, blue triangles, green
    inverted triangles, pink left triangles. The errors are smaller
    than the size of the points.}\protect\label{fig:11}
\end{figure}

\begin{figure}
  \includegraphics[width=3.5in]{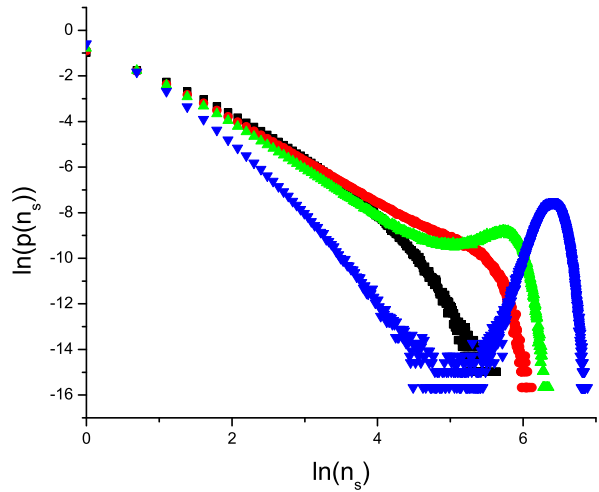}
  \caption{(Color online) The distribution of cluster sizes on the 3d
    cubic lattice for $L=12$ at inverse temperatures $\beta=0.295,
    0.335, 0.370, 0.500$ (black squares, red circles, green triangles,
    and blue inverted triangles). The errors are of the size of the
    points except for very small $p$ where they can be judged by the
    scatter (note the logarithmic scales).}\protect\label{fig:12}
\end{figure}

%COMMENT 4 HERE 

With so many clusters the average cluster at each time step must be
rather small. We measure the number of damaged sites $D\,L^d$ and the
number of clusters $n_c$, giving us the mean cluster size at each time
step. The normalized time average (and then the sample average, though
the sample variation is very small), is then $s(\beta,L)=\langle
D\,L^d/n_c\rangle$.  In Figure~\ref{fig:11} we plot $s(\beta,L)$
versus $\beta$, again for $d=3$. The curve clearly suggests a positive
right limit at $\beta_{ds}$. To estimate this limit we fit a simple
expression $a_0+a_1\,\exp(a_2\,\beta)$ to the points resulting in
$s(\beta_{ds})=\lim_{\beta\to\beta_{ds}^+}\lim_{L\to\infty}s(\beta,L)=3.46(1)$.
We estimated the right limits corresponding to $d=2,3,4,5$ to be
respectively $9.00(2)$, $3.46(1)$, $2.76(1)$ and $1.95(1)$ sites at
$\beta_{ds}$.  One could alternatively define $s$ as the average
damage divided by the average number of clusters. This is not strictly
the same as our present definition (mean ratio versus ratio of the
means) but the difference is of course vanishingly small here. For
example, the maximum in Figure~\ref{fig:10} at $\beta_{\max}=0.313$
gives $n_c\sim 0.0227\,L^3$ and the number of damaged sites, see
Figure~\ref{fig:6}, is $0.139(1)\,L^3$. Hence the average cluster size
is roughly $6.1$ which matches Figure~\ref{fig:11}.

%\begin{figure}[!ht]
%  \includegraphics[width=3.5in]{cube3d-p624406-meanclustersizevbeta.pdf}
 % \caption{(Color online) $s(\beta)$ versus $\beta$ for $L=12, 16, 24,
 %   32, 48$ (red, green, blue, purple, orange). Black dot at
 %   $\beta_{ds}$.}  \protect\label{fig:csize}
%\end{figure}

We end this section with a final remark concerning the distribution of
cluster sizes. At each time step we here pick one cluster uniformly at
random and measure its size. In Figure~\ref{fig:12} we show a set of
size distributions (or density functions) as $\log \Pr$ vs $\log s$
for a range of $\beta$ for $d=3, L=12$. They all have a high value at
size $1$ (many isolated sites) and then drop quickly. They can be
expected to behave like this since the mean size is between $4$ and
$70$ in this temperature range.  However, for $\beta>\beta_{\max}$ (or
thereabout) the distributions show a second, and rather wide, maximum
located in the neighborhood of the total damage $D(\beta,L)$. This can
be understood as follows. Suppose for the sake of argument that the
damaged sites are distributed at random in space. (This is only
approximate as correlations between damage sites should be allowed
for). Then when $D(\beta)$ exceeds the site percolation concentration
there will exist a single "giant" percolating cluster of damaged sites
together with residual small clusters. As $D(\beta)$ increases further
the percolating cluster will contain almost all the damaged sites. On
this criterion the peak should appear in the distribution at $D(\beta)
\sim 0.59, 0.31, 0.20$ and $0.14$ in dimensions $2,3,4$ and $5$
respectively which gives an indication in rough agreement with the
data.  Indeed in the case of the square the peak never appears which
is consistent with the fact that $D(\beta)$ never approaches
$0.59$. For the dimensions where the peak does appear the global
parameters, in particular $D(\beta)$, increase smoothly with $\beta$
and show no sign of any critical behavior as the giant cluster forms.

%\begin{figure}[!ht]
 % \includegraphics[width=3.5in]{cube3d12-p624406-sizedist.pdf}
  %\caption{(Color online) Distribution of cluster sizes (probability
   % versus size) for the simple cubic lattice with $L=12$ at
    %$\beta=0.295$,$0.325$,$0.355$,$0.385$, $0.415$, $0.445$,
 %   $0.475$. Distributions farther to the right corresponds to larger
  %  $\beta$.}  \protect\label{fig:cdist3d}
%\end{figure}

%\begin{figure}[!ht]
%  \includegraphics[width=3.5in]{square128-p750000-sizedist.pdf}
%  \caption{(Color online) Distribution of cluster sizes (probability
%    versus size) for the simple square lattice with $L=128$ at
%    $\beta=0.59$,$0.61$,$0.63$,$0.65$, $0.67$, $0.69$.  Distributions
%    farther to the right corresponds to larger $\beta$.}
%  \protect\label{fig:cdist2d}
%\end{figure}

\section{Conclusion}

The exact values of the coordinates of the intersection point where
the Fortuin-Kasteleyn transition line crosses the Nishimori line can
be derived for an RBIM from the analytic condition that satisfied
bonds are uncorrelated on the NL \cite{nishimori:86}. On any lattice
this leads to the exact expressions $p_{NL,FK}=(1+P_{c})/2$ and
$T_{NL,FK} =2/\ln[(1+P_{c})/(1-P_{c})]$
\cite{yamaguchi:10,campbell:94} where $P_{c}$ is the standard random
bond percolation concentration for the particular lattice. For
lattices in dimension two (and probably in higher dimensions also) the
uncorrelated bond condition remains a very good approximation at the
FK transition temperature over a wide strong-disorder region spanning
$p=1/2$.

In pure ferromagnets $T_{FK}(p=1) = T_{ds}(p=1)= T_{c}(p=1)$.  The
conjectured equivalence for the RBIM between the FK transition
temperature and the heat bath damage spreading temperature ,
$T_{FK}(p) \sim T_{ds}(p)$
\cite{dearcangelis:91,campbell:94,yamaguchi:10}, separating an
exponential from a chaotic dynamic regime has been tested at
$p=p_{NL,FK}$ on simple cubic lattices in dimensions two, three, four
and five. It holds to within $0.1\%$ in dimension three, to within
$2\%$ in dimension four, and to $3\%$ in dimension five.  The
equivalence appears always to be a good approximation. We have no
explanation to propose for the quasi-equality in the case of dimension
three.

For the square and triangle lattices the difference is larger : $5\%$
and $16\%$ respectively.

The FK transition in the strong-disorder regime close to $p=1/2$ can
thus be said to be well understood. However the basic physical
condition determining the the damage spreading transition temperature,
which plays an important role in limiting perfect equilibration in
RBIMs, and its proximity to the FK transition, remains unclear.

\section{Acknowledgements}
We are very grateful for enlightening remarks by H. Nishimori,
C. Yamaguchi, R. Ziff and W. Krauth.  The computations were performed
on resources provided by the Swedish National Infrastructure for
Computing (SNIC) at High Performance Computing Center North (HPC2N).

%This research was conducted using
%the resources of High Performance Computing Center North (HPC2N).

%Figure 1  square all(T) OK

%Figure 2 triangle all (T) OK

%Figure 3 triangle FKr(p,L) OK

%Figure 4 triangle FKr(p,L)zoom OK

%Figure  5 extrapolation $\beta_{ds}$ triangle OK
%Figure 6 extrapolation $\beta_{ds}$ 3d cube  OK

%Figure 7 $T_ds (p) 3d L32 OK$

%fiure 8 same 4d

%Figure 9 snapshot  OK

%Figure 10 number of clusters cnum  3d OK

%Figure 11 csize 3d

%Figure 12 cdist3d

\end{document}